\documentclass[jcp,aip,10pt,floatfix,twocolumn,showpacs,superscriptaddress]{revtex4-1} 
\usepackage{color,soul} 
\usepackage{mathtools}
\usepackage{float}
\usepackage{graphicx}
\usepackage{hyperref}
\usepackage{amsmath, amsthm, amssymb}
\usepackage{multirow} 
\usepackage[normalem]{ulem}
\begin{document} 
\title{ Confined crowded polymers near attractive surfaces} 
\author{Kamal Tripathi} 
\email{kamalt@imsc.res.in}
\affiliation{The Institute of Mathematical Sciences, C.I.T. Campus, Taramani, Chennai 600113, India}
\affiliation{Homi Bhabha National Institute, Training School Complex, Anushakti Nagar, Mumbai 400094, India}

\author{Gautam I. Menon} 
\email{menon@imsc.res.in} 
\affiliation{The Institute of Mathematical Sciences, C.I.T. Campus, Taramani, Chennai 600113, India}
\affiliation{Homi Bhabha National Institute, Training School Complex, Anushakti Nagar, Mumbai 400094, India}
\affiliation{Department of Physics, Ashoka University, Plot No. 2, Rajiv Gandhi Education City, National Capital Region, Sonepat 131029, India}

\author{Satyavani Vemparala} 
\email{vani@imsc.res.in} 
\affiliation{The Institute of Mathematical Sciences, C.I.T. Campus, Taramani, Chennai 600113, India}
\affiliation{Homi Bhabha National Institute, Training School Complex, Anushakti Nagar, Mumbai 400094, India}

\begin{abstract}
We present results from  molecular dynamics simulations of a spherically confined neutral polymer in the presence of crowding particles, studying polymer shapes and conformations as a function of the strength of the attraction to the confining wall, solvent quality and the density of crowders.  The conformations of the polymer under good solvent conditions are weakly dependent on crowder particle density, even when the polymer  is strongly confined. In contrast, under poor solvent conditions, when the polymer assumes a collapsed conformation when unconfined, it can exhibit transitions to two different adsorbed phases, when either the interaction with the wall or the density of crowder particles is changed. One such transition involves a desorbed collapsed phase change to an adsorbed extended phase as the attraction of the polymer towards the confining wall is increased. Such an adsorbed extended phase can exhibit a second transition to an ordered adsorbed collapsed phase as the crowder particle density is increased. The ordered adsorbed collapsed phase of the polymer differs significantly in its structure from the desorbed collapsed phase. We revisit the earlier understanding of the adsorption of confined polymers on attractive surfaces in the light of our results.
\end{abstract}

\maketitle
\section{Introduction} 
For an unconfined neutral polymer, the nature of the solvent determines whether the polymer adopts, on average,  an extended or a collapsed conformation. If confined,  the polymer experiences a loss of conformational entropy.  This reduces allowed configurations to a subset that depends on the shape and other properties of the confining volume~\cite{ganai2014chromosome, jun2006entropy, jun2010entropy, kang2015confinement, Tark-Dame839, jeon2016effects, fovsnarivc2013monte, ruggiero2018confinement}.  

A number of polymers in biological contexts encounter varying degrees of confinement. The approximately $2m$ of DNA in the nucleus of eukaryotic cells must be restricted to a nucleus that is $\sim 10\mu m$ in radius, while the packaging of viral DNA into sub-micron-sized protein capsids is often dense enough to induce local crystallinity ~\cite{boyle2001spatial, misteli2007beyond, lieberman2009comprehensive}. Cargo transported along axons by molecular motors typically consists of specific protein molecules encapsulated in vesicles of diameter $30 - 80 nm$, comparable in dimension to the larger peptide neurotransmitters that such vesicles must accommodate \cite{purves2001organization}.  

Biological polymers {\emph {in vivo},} in addition to being confined in their natural contexts, also inhabit highly crowded environments. The presence of crowders can affect the compaction and higher-order organization of single biopolymer, as well as promote aggregation of such polymers in solutions ~\cite{ha2015polymers, zimmerman1993macromolecular, phillips2012physical, zhou2013influence, feig2017crowding, lim2014polymer,ellis2001macromolecular}. At high crowder concentration, neutral polymers have been shown to undergo a continuous extended-to-collapsed transition \cite{shendruk2015simulating, van1999effects}. Incorporating attractive interactions between crowder particles and macromolecular polymers leads to the formation of complex aggregates which can be observed directly \cite{zhou2008macromolecular}.  Although repulsive interactions do not promote the formation of complexes, they can affect reaction rates and conformations via the Asakura Oosawa depletion interaction \cite{asakura1954interaction}. Experiments show that small molecules such as polyethylene glycol (PEG) can condense DNA \cite{vasilevskaya1995collapse}.  

Athough DNA is confined and crowded within the nucleus, it is not structureless and the interplay between polymer shape, confinement and crowding can potentially accentuate certain aspects of such a structure while attenuating others. Within the cell, the presence of a heterogeneous mixture of proteins, organelles, water and ions can influence the loss of conformational entropy of the biopolymers present in ways that are particularly hard to predict\cite{zimmerman1993macromolecular, minton2005models, uversky2002accelerated, cheung2005molecular, bokvist2007misfolding}.  As regards the effects of confining polymers by surfaces, understanding polymer adsorption onto surfaces as solvent conditions are varied is relevant both to the efficiency of biosensors \cite{yang2015carbon} as well as to the formulation of protein resistant coatings~\cite{ionov2009protein, haraguchi2014highly}. The ability to accurately describe the phase behavior of polymers in confined, crowded  regimes is thus central to an improved understanding of a number of biological processes~\cite{strulson2014molecular, onuchic2004theory, thirumalai2005rna, iborra2007can}.  

Using atomistic simulations to study the structural and dynamical properties of realistic bio-polymers is computationally expensive, even without accounting for crowders. Thus, many studies use a coarse-grained, neutral polymer description of such biopolymers, even though biologically relevant polymers such as DNA are typically weakly charged in solution. (The screening of the charges through counterions, salt and the presence of other charged species around the biological polymer of relevance should render the electrostatic interactions effectively short-ranged\cite{nelson2004biological}.) A number of coarse-grained simulation studies of neutral polymers in the presence of crowders have been performed earlier \cite{reddy2006implicit, jeon2016effects, kang2015confinement, shin2015kinetics, lim2016depletion, lim2016influence, davis2018influence, louis2002polymer}.  Several  other studies have examined the interaction of both neutral and charged polymers with surfaces\cite{rajesh2002adsorption, reddy2010solvent, bachmann2005conformational, plascak2017solvent, martins2018adsorption}. 

The interplay between chain entropy, monomer-monomer interactions and monomer-surface interactions should determine the conformational landscape of polymers near surfaces.  There appear to be four dominant ``phases'' that describe the behaviour of neutral polymers near attractive walls and under different solvent conditions: the desorbed-extended (DE), desorbed-collapsed (DC), adsorbed-extended (AE) and adsorbed-collapsed (AC) phases~\cite{rajesh2002adsorption, bachmann2005conformational, arkin2012ground, krawczyk2005layering}.  Using phenomenological arguments, Rajesh {\it et al.} \cite{rajesh2002adsorption} showed that for low values of attraction between the monomers and the surface as well as between the monomers, the polymer adopts a desorbed-extended conformation (DE). Upon increasing the monomer-monomer attraction, a desorbed-collapsed (DC) phase results.  If the interaction strength between monomer and surface is further increased, the polymer can assume either an adsorbed-extended (AE) or an adsorbed-collapsed (AC) conformation, depending on the monomer-monomer interaction. These authors also predicted a ``surface attached globule'' state (SAG), in which the number of contacts between the collapsed conformations of the polymer and the surface is much less than that in AC phase. 

Monte-Carlo simulation studies of neutral and spherically confined polymers suggest that at low temperatures, where enthalpic effects dominate, the polymer adsorbs on the wall of the sphere,  tending to form layer-like structures~\cite{arkin2012structural, bachmann2005conformational}.  At higher temperatures, where entropic effects dominate, the polymer was found to desorb, assuming an extended conformation (DE). The effect of solvent conditions on polymer conformations near the surface have also been studied using lattice models of a grafted polymer on a surface \cite{plascak2017solvent}. Under good solvent conditions, a grafted polymer on a flat surface is adsorbed on the surface at low temperatures, while  it assumes a desorbed conformation at higher temperatures. Under poor solvent conditions and for low temperatures, the polymer takes a globular adsorbed conformation, while for higher temperatures, the polymer desorbs from the surface. 

Multi-canonical Monte Carlo simulations suggest a possible phase diagram consolidating the observed behaviour~\cite{arkin2012structural}. In this phase diagram, for low temperatures and for a small value of monomer-surface interaction energy, the polymer assumes a desorbed collapsed conformation.  With increasing surface interaction, the polymer undergoes a transition from an amorphous globular conformation to a more layered internal structure. The layering, in those simulations, covered regimes ranging between a  $4-$layered adsorbed structure to  an adsorbed monolayer. 

Although various aspects of confinement and crowding in relation to polymer conformation near attractive surfaces have been studied before, we know of no studies that examine the interplay of all three parameters on an equal footing. In addition,  we note  that the simulations, especially those for attractive surfaces, have  largely been performed on polymers of relatively short chain length. The  layered structures observed in these regimes in previous work could thus be an artefact of the small polymer size.  Different structures could possibly be stabilized, or the boundaries between the states proposed earlier altered, when long chain polymers,  as opposed to short ones,  are adsorbed at a surface. In addition, confining polymers in three dimensions and adding crowders to the system adds  further dimensions of complexity, but represent a scenario that is more relevant to biophysical situations. 

To examine these questions, this paper explores how confinement, crowder density, solvent conditions and surface interaction combine to influence polymer conformations, using simulations of  a simple model system.  Our work extends previous results through the incorporation of the effects of crowder density, specifically, in the confined case with wall interaction. We point out that solvent quality and crowder density complement each other in determining configurations, and that the resultant effects of these are most prominent in the poor solvent case. We characterize the ``crumpling'' of polymer conformations under the combination of high crowder density and poor solvent conditions. Although the parameter space is large, our results provide an understanding for the nature of polymer conformations in each of these different regimes, suggesting physical arguments for why they should be stabilized.

This paper is organized as follows: in Section II,  we discuss the procedures by which we  set up our systems, describing the simulation protocols employed and the methods used for our calculation. Our results are presented in section III. This is followed by a discussion section, Section IV  that summarizes the conclusions we draw from our study and provides suggestions for future work.
 
\section{Methods \label{sec:model}} 
We study, using molecular dynamics simulations, a single, long, self-avoiding polymer chain of $400$ monomers, confined to the interior of a hollow sphere. The simulated volume also contains crowders {\it i.e.} particles that interact non-specifically with the monomers constituting the chain as well as with themselves. In our simulations, we vary solvent quality across the extreme limits of good and poor solvents. There are no other explicit solvent particles, aside from the crowder particles. In previous work~\cite{kang2015effects}, a parameter $\lambda$ was defined to compare the relative sizes of the crowder particle and the polymer ($\lambda=R_g^0/\sigma_c$, where $R_g^0$ is the radius of gyration of the polymer with no crowder particles present and $\sigma_c$ is the size of the crowder particle). Following this definition, the sizes chosen in our present simulations correspond to $\lambda >> 1$, as the monomer and the crowder particle have the same size. Also, note that in the earlier simulations, hard-sphere potentials were used to describe monomer-monomer interactions, unlike the interaction potentials used in this study, Hence, an exact mapping between $\lambda$ values is not possible and the definition should only be treated as an approximation.

We define the effective crowder density as $\phi_c = N_c v_c/ V$, with $N_c$ being the number of crowders, $v_c = 4/3 \pi \sigma_c^3$ the volume of each crowder particle, $\sigma_c$  the effective radius of the crowder, and $V$  the effective volume of the spherical confining region. We vary $\phi_c$  between $0.035$ to $0.435$ in steps of $0.05$, by changing the number of crowder particles between $950$ and $11750$. The range of crowder densities considered in this work is similar to the range used in previous work~\cite{kang2015effects, kang2015confinement, palit2017effect}.  We also perform simulations of a polymer in a good solvent placed in a periodic box, thus mimicking the unconfined case, so as to compare our results with results for the confined case.  The crowder density range explored for such simulations is the same as for the confined case. 

Pairs of all non-bonded particles (monomers and crowders) interact through the van der Waals interactions, modeled through a truncated and shifted Lennard Jones (LJ) 6-12 potential with a cutoff of $r_c$ which is given by
\begin{equation}
V (r_{ij}) = V^{LJ}(r_{ij}) - V^{LJ}(r_c),
\label{eq:WCA}
\end{equation}
where
\begin{equation}
V ^{LJ}(r_{ij}) =
4 \epsilon_{ij}\left[ \left(\dfrac{\sigma_{ij}}{r_{ij}}\right)^{12} - \left(\dfrac{\sigma_{ij}}{r_{ij}}\right)^{6} \right], 
\label{eq:LJ}
\end{equation}
and $V^{LJ}(r_c)$ is the shift term to make the potential zero at the cutoff $r_c$.
Here $i, j = m, c$ refers to the monomers and crowder particles respectively, and $r_{ij}$ is the distance between two particles, (By definition, $\epsilon_{ij}$ is a symmetric matrix and the quantities $\epsilon_{ij}$ and $\sigma_{ij}$ define the corresponding interaction parameters.) The size of the monomer and the crowder particles is set to be $1.0$, in our reduced units. We model different solvent conditions by choosing different values for the cutoff distance $r_c$ for the interactions among the monomers. The good and poor solvent conditions are reproduced via interactions among the monomers with different $r_c$ (see Table  ~T1 of Supplementary Information). The nature of the interactions among crowder particles and between crowders and monomers is a repulsive, soft-core interaction ($r_c=2^{1/6}$). The interaction of the monomers with the wall is also described by a Lennard-Jones potential and both repulsive and attractive wall interactions between monomers and wall are considered. The interaction of the crowder particles with the wall is set to be repulsive regardless of solvent condition. The strength and sign of the interaction of the polymer with the wall can be varied, so that the full range between repulsive and attractive wall strengths is accessed. The parameters used in this study are given in Table ~T1 of Supplementary Information.

The polymer chain connectivity is modelled via a harmonic potential
\begin{equation}
V_{ij} ^{bond}(r_{ij}) = \frac{1}{2} k_{bond} (r_{ij}-r_0)^2.
\label{eq:harmonic-pot}
\end{equation}
where the bond length $r_0$ for the polymer is set to be $1.122$.
The equation of motion is integrated for $10^8$ steps using a velocity-Verlet algorithm. The step size is taken to be $\delta t = 0.001\tau$, where $\tau = \sigma \sqrt{m/\epsilon}$, $m$, $\sigma$ and  $\epsilon$ are units of mass, length and energy respectively. All simulations are performed under constant volume and temperature (T) conditions $(T = 1.0)$ using a Nose-Hoover thermostat. The MD implementation is from the LAMMPS \cite{plimpton1995fast} software package. All visual image generation and analysis was performed  using scripts developed in the VMD package ~\cite{HUMP96}. The system's initial configuration is constructed using the Pizza-py toolkit ~\cite{plimpton1995fast}. A harmonic wall interaction was initially used for $10^4$ steps to stabilise the polymer inside the confining surface (of radius $R = 15.0$). With this starting point, several systems with the desired wall interaction potential and parameters were generated.  

Different shape parameters are calculated to assess the size and shape of the polymer, as detailed below.  The radius of gyration is defined as
\begin{equation}
R_g ^2 = \frac{1}{2N^2}~\sum_{i=1}^N \sum_{j=1}^N |({\vec r}_i - {\vec r}_j)|^2,
\label{eq:rgt}
\end{equation}
where $r_i$  and $r_j$ are the position vector of $i^{th}$ and $j^{th}$ particle respectively.

Additional shape parameters can be defined using  the definition of the  gyration tensor,
\begin{equation}
S_{mn} = \frac{1}{2N^2}\sum_{i = 1} ^{N} \sum_{j = 1} ^{N} \left( r_i ^m - r_j ^m\right) \left( r_i ^n - r_j ^n\right), 
\end{equation}
where $r_{i}^m$ is the $m^{th}$ cartesian coordinate of the position vector $r_i$ of the   $i^{th}$ particle

The gyration tensor can be diagonalised, yielding three eigenvalues $\lambda_1, \lambda_2, \lambda_3$ (where we order  $\lambda_1< \lambda_2< \lambda_3$ ). The asphericity $b$, which describes the deviation of the average shape of the polymer from a sphere, is defined as,
\begin{equation}
b = 1 - \frac{1}{2\lambda_3} (\lambda_1 + \lambda_2).
\end{equation}

The local structure of polymer conformations can be represented through  contact maps. If any two monomers, say monomer $i$ and monomer $j$,  approach each other to within a distance of $2 \sigma$, this is counted as a contact between $i$ and $j$. Contact maps are calculated over a production run ($5 \times 10^6\tau$). An average value for the number of contacts is then computed and associated to elements of a two dimensional matrix indexed by monomer labels. 

For the confined, adsorbed case, we count the average fraction of monomers adsorbed on the surface. We also measure a height function that quantifies the height of the adsorbed configuration relative to the confining sphere surface. To calculate the height of the polymer stacks, we calculate the distance of each monomer from the centre of the confining sphere. We then subtract this quantity from the radius of the confinement. The largest number thus obtained, after averaging over a large number of configurations,  is a measure of the height of the polymer stack, and is termed $H$. 

To understand the effect of the curvature of the confinement on the conformation of the polymer, we  performed additional simulations with a larger sphere of radius $R = 30.0$,  for the same crowder densities and potential parameters. In all cases, we perform a large number of simulations starting from different initial conditions and with the same parameters to ensure good statistical averaging. 


\section{Results}
\subsection{Effect of confinement and crowders on polymer conformations in a good solvent}
To understand the impact of crowder particle density and confinement on polymer conformation, we simulated the neutral polymer within a confining sphere with repulsive walls, and under good solvent conditions. A control simulation with periodic boundary conditions was also performed to compare  results for confined and unconfined polymers.  
\begin{figure}
\includegraphics[width = 1.0\columnwidth]{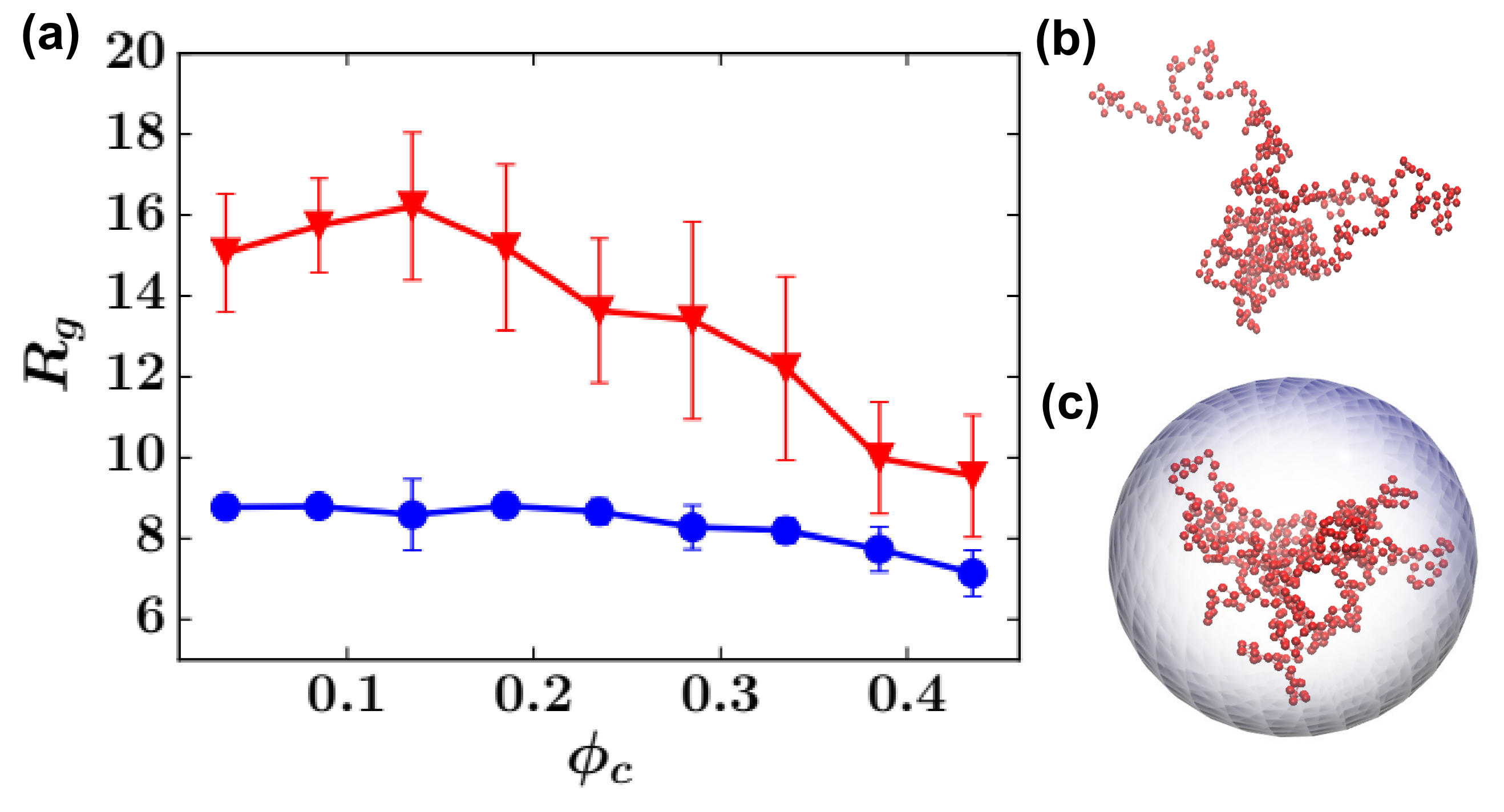}
\caption{ \textbf{(a)} The radius of gyration of the polymer in a good solvent condition under confinement with repulsive walls (blue) and in the absence of confinement (red) at different crowder densities. Each data point for unconfined and confined polymer is averaged over 10 and 5 simulations respectively. The corresponding error bars are also shown in the figure which represent standard deviation of the data. \textbf{(b,c)} Snapshots of the polymer at the highest crowder density are shown for periodic boundary conditions and spherical confinement.}
\label{fig:gsrw}
\end{figure}
We measured the radius of gyration $R_g$ of the polymer, under both confining and non-confining conditions, varying crowder densities. Our results are shown in Figure~\ref{fig:gsrw}. The radius of the confining sphere was chosen based on the typical $R_g$  value of the polymer in the unconfined case ensuring that even at the smallest crowder particle densities, the $R_g$ value of the polymer in the confined sphere is significantly lower than that in the unconfined case. 

Earlier studies~\cite{kang2015confinement,marenz2012simple,hsu2005coil,shendruk2014coarse,cacciuto2006self,das2010effect}, both theoretical and computational, have shown that the primary effect of confinement is to reduce the conformational space available to the polymer. This can be seen in our simulations at very low crowder density. As the crowder density is increased, the conformations of confined and unconfined polymer vary differently with the crowder densities. As seen in earlier work ~\cite{kang2015effects, woodward2010depletion, woodward2012many,bolhuis2002influence,bolhuis2003colloid, wang2014exact} for the unconfined case, $R_g$ values decrease,  indicating an evolution into a more compact structure, as the crowder density is increased. Figure~S1 (Supplementary Information), shows the relative change with respect to the lowest crowder density considered in this study. 

From both Figure~\ref{fig:gsrw} and Figure~S1, it can be seen that for the highest crowder density simulated here, the reduction of $R_g$ is significant, as also seen in other studies~\cite{mao1995depletion,kim2015polymer,shendruk2014coarse,jeon2016effects}. This is a consequence of depletion forces due to the crowders.  However, note that the decrease is not as dramatic as that seen in Ref~\cite{kang2015effects}, where confinement effects are not considered. In addition, the definitions of $\lambda$ differ, making a direct comparison difficult. Finally, the polymers used in the present study are significantly longer than those used in previous work. In Figure~S2, we compare the dependence of the radius of gyration for $N=50$ with the $N=400$ case (for $\epsilon_{mc} = 5.0$) under good solvent conditions, which shows the dependence of $R_g$ on the degree of polymerization. For short chains, the confinement effects are minimal compared to the crowder effects, while the confinement plays a substantially bigger role for longer chain polymers. This can be understood in terms of the ratio of the average polymer size and the confinement radius.

For the confined case, the $R_g$ values of the polymer are relatively insensitive to the crowder densities when compared to the unconfined polymer, with only a slight reduction at highest values of crowder densities. This suggests that under confinement, the polymer conformation is already somewhat compact  and that the addition of crowder particles thus does not change the global conformation significantly. At the highest crowder densities considered in the present study, $R_g$ values of unconfined and confined polymer are comparable. However, introducing an attraction between monomers, modelling poor solvent conditions, can affect the conformational landscape of the polymer. This will be explored in later  sections.

\subsection{Confined polymers in a good solvent with attractive walls }
\begin{figure}
\centering
\includegraphics[width=\columnwidth]{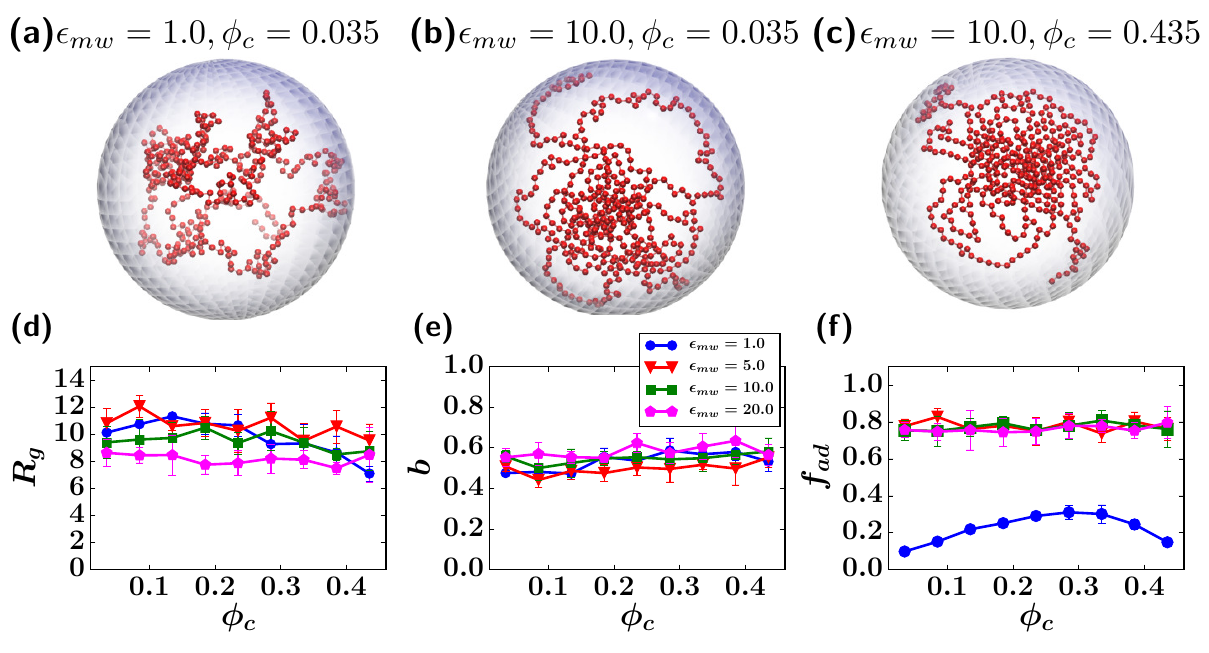}
\caption{Good solvent condition: The snapshots of the polymer conformation for \textbf{(a)} the wall attraction $\epsilon_{mw} = 1.0$, the crowder density $\phi_c = 0.035$, \textbf{(b)} $\epsilon_{mw} = 10.0, \phi_c = 0.035$, \textbf{(c)} $\epsilon_{mw} = 10.0, \phi_c = 0.435$.  The variation of \textbf{(d)} radius of gyration $R_g$, \textbf{(e)} asphericity $b$ and \textbf{(f)} the fraction of adsorbed monomers $f_{ad}$ with $\phi_c$ for different $\epsilon_{mw}$. Here blue circle, red inverted triangle, green square and magenta pentagon refers to $\epsilon_{mw} = 1.0, 5.0, 10.0, 20.0$ respectively.}
\label{fig:gs}
\end{figure}

In this section, we explore the conformations of a long polymer  interacting with an attractive wall and  under good solvent conditions. Representative snapshots of the systems for different wall attraction strengths and crowder densities are shown in Figure~\ref{fig:gs}. In the previous section, we showed that in the confined case and for repulsive walls, the global conformation of the polymer varies little with crowder particle density. Figure~\ref{fig:gs}(a) shows that for attractive wall interactions,  there is only a marginal change in the overall conformation of the polymer, provided a good solvent condition is maintained.

To quantify this visual observation further, we computed two shape parameters for the polymer: (i) asphericity ($b$) and (ii) radius of gyration $R_g$, both as a function of crowder density. In addition, we also compute the number of adsorbed monomers on to the confining surface also as a function of crowder density. The variation of $R_g$ values plotted against crowder density for  a range of  attractive wall strengths is shown in Figure~\ref{fig:gs}(d). This figure shows  that the conformation of the polymers is largely insensitive to the crowder density, even at the highest crowder densities we consider. This is in contrast to the repulsive wall case discussed in the previous section. 

This result suggests that the attractive wall interactions dominate the conformational landscape of the polymer when it is confined  under good solvent conditions. A small reduction in $R_g$ values upon increasing the crowder density is only seen when the attractive wall strength is small, for $\epsilon_{mw}=1.0$ in the Figure~\ref{fig:gs}.

These results are mirrored in the plots of asphericity as a function of crowder density shown in Figure~\ref{fig:gs}(e). At the largest values for the attractive wall interaction, $b$ values are relatively insensitive to crowder densities, but appear to exhibit somewhat unusual and non-monotonic behaviour for intermediate values of crowder density and wall interaction. To further understand the adsorption of the polymer on to the confining wall surface,  we measure the number of adsorbed monomers on the surface ($f_{ad}$)  for different wall attraction strengths as a function of crowder densities. We consider a  monomer to be adsorbed on the surface if it is within $1.5 \sigma$ of the wall. Each point in the plot represents an average of $5$ initial conditions and is averaged over $5 \times 10^6$ time steps.

For the lowest values of wall attraction strength and for a small number of crowder particles, the number of adsorbed monomers is very small, suggesting a largely desorbed-extended (DE) configuration of the polymer. However, as the crowder density is increased, there is an increase in the number of adsorbed monomers on the surface. This observation is consistent with our results for the $R_g$ and $b$ values of the polymer for low attractive wall strength and highest crowder density, where we noted that the polymer tends  to have a more compact structure and the effects of crowder particles dominate. However, as the wall attraction strength is increased, i.e. for $\epsilon_{mw}>1.0$, there is a substantial increase in the number of adsorbed monomers, as indicated by the adsorption of nearly $80\%$ of monomers, in Figure~\ref{fig:gs}(f). This  does not vary much with crowder density and the polymer remains in  an adsorbed-extended  (AE) state under good solvent conditions for high attractive wall strengths. 

We believe that the non-monotonic behaviour of the asphericity in Figure~\ref{fig:gs}(e) is likely due to the fact that the polymer shape is considerably distorted by its interactions with the surface and its spreading. At low values of wall interaction, both  Figure~\ref{fig:gs}(d) and  Figure~\ref{fig:gs}(e) suggest that the polymer is largely detached from the wall. At the highest values of wall interaction, the polymer is largely adhered to the wall. For values of the wall interaction that lie in-between these extremes, although most monomers are adhered to the wall on average, excursions from it are not penalized as much, especially at low crowder concentrations. As the polymer shifts between a largely three-dimensional to a largely two-dimensional conformation, the nature of the depletion interaction induced by the crowders can exhibit a non-trivial dependence on crowder density, which we suggest may be responsible for the complex behaviour we see.

Note that the results presented in this section are for  $\epsilon_{mc} = 0.5$ and good solvent conditions, different from the poor solvent conditions which are the main focus of the paper (see Table T1 in Supplementary Information). However we also used a larger value of $\epsilon_{mc} = 5.0$ (a higher repulsion between monomer and crowder) in order to check whether similar results are obtained. In Figure~S3, we compute configurational properties of our system with $N = 400$ and $\epsilon_{mc} = 5.0$ in good solvent conditions. The polymer conformation and coverage near the attractive surface are seen to be a result of a competition between the crowder effects and the attraction of the monomers to the wall. If $\epsilon_{mc}$ is small, wall effects dominate at lower values of $\epsilon_{mw}$ . When we increase $\epsilon_{mc}$ to $5.0$, the crowder effects are slightly larger for good solvent conditions as can be seen in Figure~\ref{fig:gs} and Figure~S3. As seen in these figures, increasing $\epsilon_{mc}$ under good solvent conditions acts mainly to increase the coverage.

\subsection{Confined polymers in a poor solvent with attractive walls}
\begin{figure}
\centering
\hspace{20pt}
\includegraphics[width=\columnwidth]{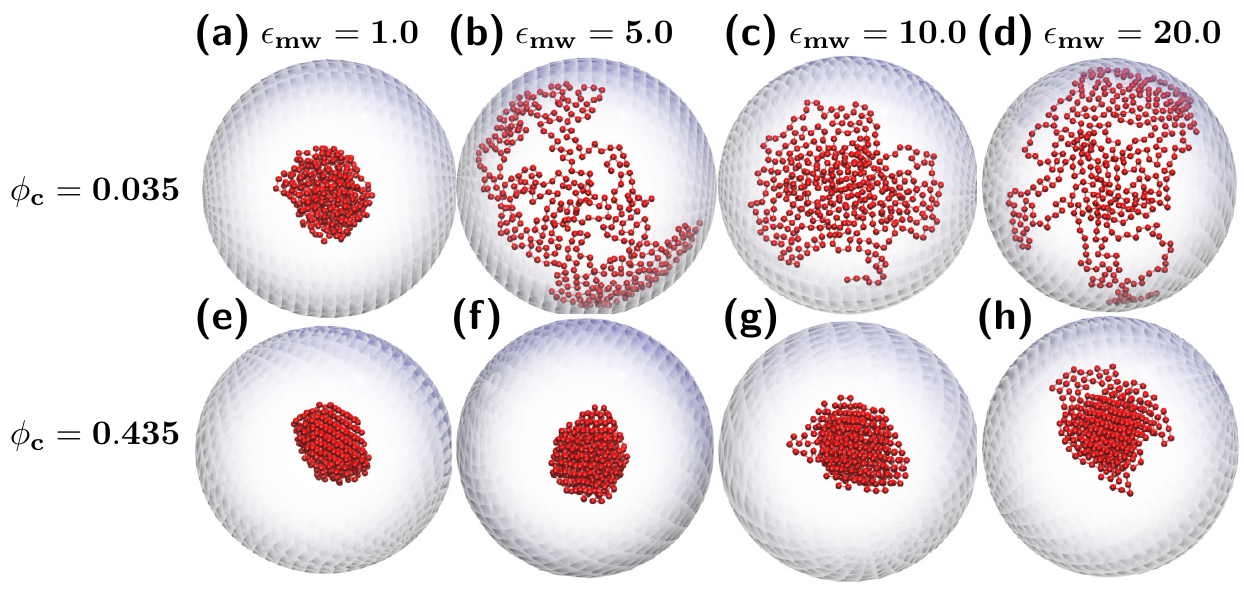}
\caption{Poor solvent: The snapshots of the polymer  for low crowder density $\phi_c = 0.035$ (a-d)  and for high density $\phi_c = 0.435$ (e-h) for different wall attractive strengths.}
\label{fig:snaps-diffE13}
\end{figure}
A neutral, unconfined polymer is collapsed under poor solvent conditions. In this section, we explore confined polymer conformations under poor solvent conditions and near an attractive wall. We vary the strength of the interaction with the wall as well as the crowder particle densities, mapping out the qualitative phases obtained with these parameters.  

For low crowder densities, $\phi_c = 0.035$, and for a small monomer-wall attractive strength $\epsilon_{mw} = 1.0$, the polymer assumes a globular conformation on average. This can be seen in Figure \ref{fig:snaps-diffE13}(a). If we now increase the attraction towards the wall, keeping crowder densities the same, the polymer extends while remaining adsorbed on the wall surface (AE), as seen in  Figure \ref{fig:snaps-diffE13}(b-d)). However, at high crowder densities, the polymer remains in an adsorbed collapsed conformation (AC), regardless of the wall interaction strengths, as shown in Figure \ref{fig:snaps-diffE13}(e-h). This is in contrast to what was obtained for good solvents, for which the polymer remains in an AE conformation at sufficiently large wall interaction strength, irrespective of crowder densities. 

\begin{figure}
\centering
\hspace{20pt}
\includegraphics[width=\columnwidth]{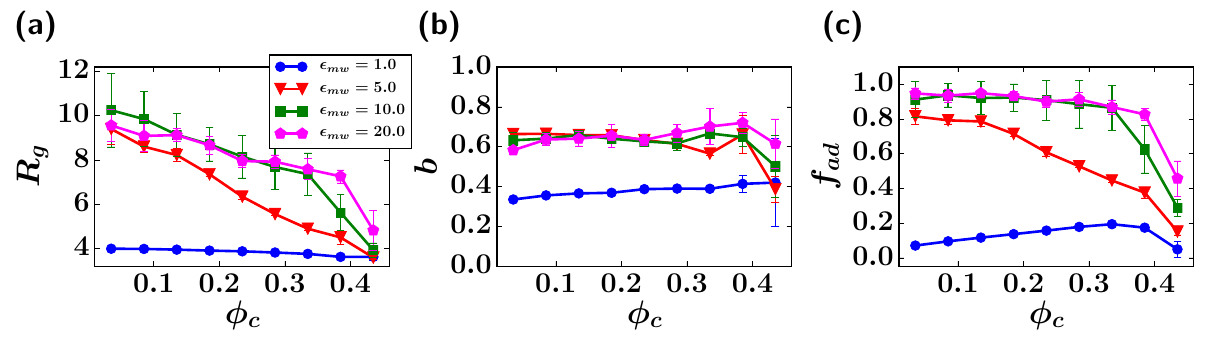}
\includegraphics[width=\columnwidth]{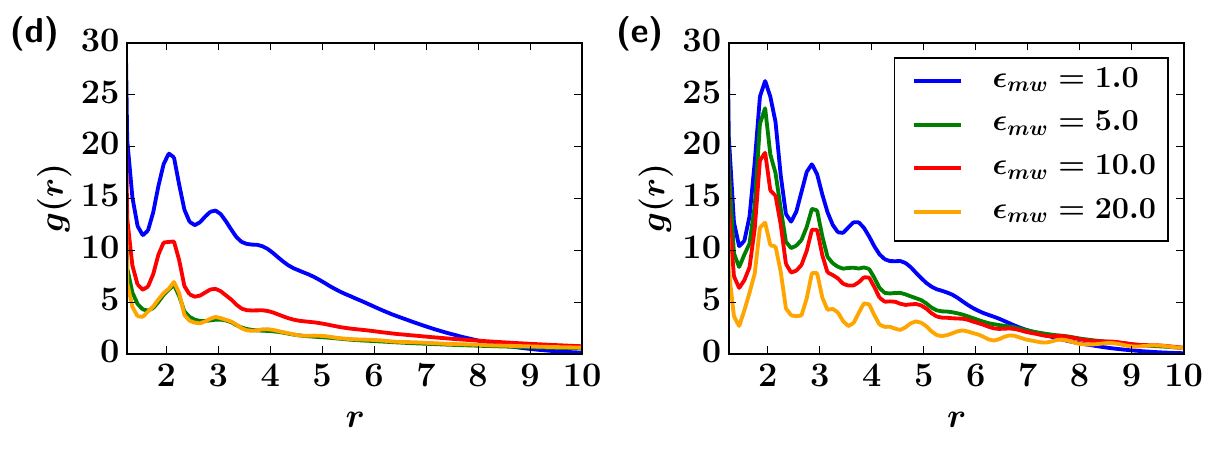}
\caption{Poor solvent: \textbf{(a,b,c)} Different shape parameters of the polymer for various attractive wall strengths as a function of crowder density. In figures (a), (b) and (c) blue circle, red inverted triangle, green square and magenta pentagon refers to $\epsilon_{mw} = 1.0, 5.0, 10.0, 20.0$ respectively. The pair radial distribution functions for \textbf{(d)} low crowder density $\phi_c = 0.035$ and \textbf{(e)} high density $\phi_c = 0.335$ for different wall attractive strength $\epsilon_{mw}$.}
\label{fig:Figure4-new-2}
\end{figure}

The coverage of the polymer onto the confining surface in poor solvent conditions differs significantly from that in good solvent conditions. This can be understood in terms of two opposing effects. The polymer wants to maximize its (attractive)  interactions with the wall by spreading out over it, but this competes with a tendency towards compaction induced by the poor solvent and exacerbated by the presence of crowder particles. This tendency is present even for low crowder densities. To gain further insight, we calculated the conformational parameters $R_g$, $b$ and $f_{ad}$, for different attractive wall strengths and as a function of crowder densities. These results are shown in Figure ~\ref{fig:Figure4-new-2}(a-c).  

For an attractive wall strength of $\epsilon_{mw}=1.0$, regardless of the crowder density, the polymer is in a collapsed conformation. At the lowest crowder density, as the attractive wall strength increases across $\epsilon_{mw}=5.0, 10.0$ and $20.0$, the $R_g$ values are increased, suggesting a transition from desorbed-collapsed (DC) to the adsorbed-extended(AE) structure of the polymer.  At these higher attractive wall strengths, the crowder density influences the polymer conformation significantly under poor solvent conditions. This is in contrast to good solvent conditions, under which the crowder density has a far smaller effect on the polymer conformation.

In addition, under poor solvent conditions, as the crowder density increases, the polymer makes a second transition from adsorbed-extended (AE) to adsorbed-collapsed (AC), as can be seen from the decrease of the $R_g$ values in  Figure ~\ref{fig:Figure4-new-2}(a) for higher crowder densities. This can also be seen in calculations of  the asphericity Figure ~\ref{fig:Figure4-new-2}(b), where, at high crowder densities, regardless of the wall attraction, the polymer assumes a compact conformation. This contrasts to the behaviour in good solvent conditions. The number of adsorbed monomers on the wall surface also show a similar decrease at high crowder densities, as shown in Figure ~\ref{fig:Figure4-new-2}(c).

Though the $R_g$ values of the polymer, under poor solvent conditions and for high attractive wall strengths, is similar to that of the polymer under good solvent conditions, the internal structure of the polymer at low and high crowder densities is very different. This is captured by radial distribution functions ($g(r)$)between monomers. Our results are shown in  Figure ~\ref{fig:Figure4-new-2}(d-e). The $g(r)$ values of the polymer under poor solvent and at high crowder density clearly show a more ordered internal structure when compared to low crowder density. This is seen in the increased number of peaks emerging in pair distribution function between monomers. The internal structure is more pronounced at low attractive wall strengths while the competing wall interactions at higher strengths lead to reduced structure. 

Consolidating these results provides us a global picture of the possible conformations of a neutral polymer in the $\phi_c-\epsilon_{mw}$ space under poor solvent conditions and the competing effects that favour different conformations. A small wall attraction leads to an adsorbed collapsed (AC) structure. Upon increasing the crowder density, the  polymer maintains this AC structure. As the attraction of the polymer towards the wall is increased, the AC state opens up in order to maximise contact with the wall. There is  then a transition towards an adsorbed-extended (AE) as well as a layered structure at low crowder density. If we now increase crowder density, the polymer tries to collapse further to avoid exposure to the crowder particles. The competition between the attractive interactions between monomer and wall and the repulsive interaction between the polymer and the crowder particles now governs polymer shape and internal structure. For intermediate $\epsilon_{mw}$ values, the crowder interaction appears to be dominant while for higher values of $\epsilon_{mw}$, the attraction wins, leading to an absence of layering even at higher crowder densities.

\subsection{Confined polymers with attractive walls: The role of crowder density}

\begin{figure}
\includegraphics[width=\columnwidth]{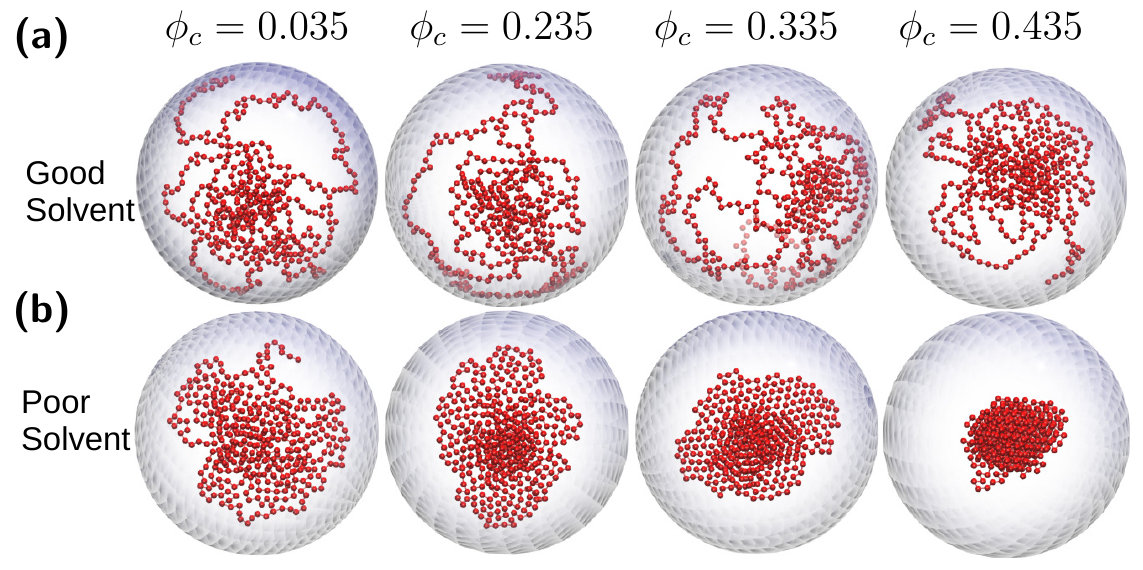}
\centering
\includegraphics[width=\columnwidth]{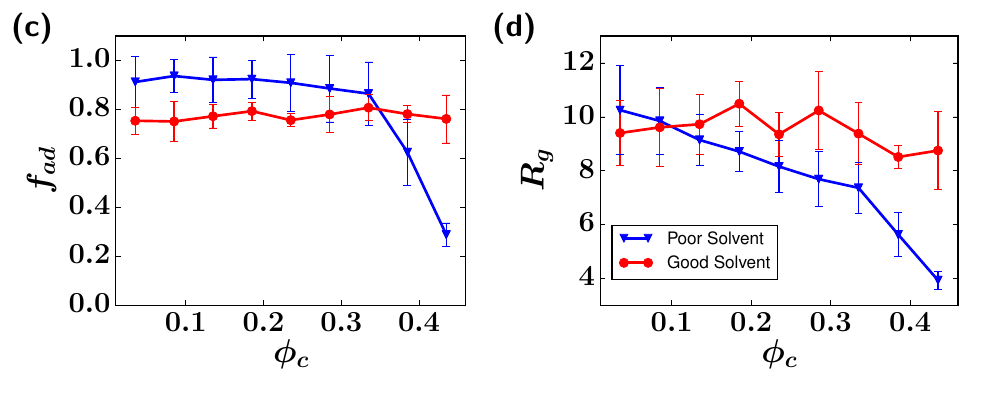}
\caption{The snapshots of the polymer \textbf{(a)} for good solvent condition \textbf{(b)} for poor solvent condition at different crowder densities $\phi_c = 0.035, 0.235, 0.335, 0.435$ for attractive wall strength $\epsilon_{mw} = 10.0$, \textbf{(c)} Fraction of adsorbed monomers on the surface of confinement as a function of crowder density in good and poor solvent conditions. \textbf{(d)} The radius of gyration of polymer as a function of crowder density for $\epsilon_{mw} = 10.0$.}
\label{fig:gsvsbs-phi}
\end{figure}

In this section, we compare the conformations and internal structure of the polymer as a function of crowder densities. Some conformations of the polymer, along with $R_g$ and $f_{ad}$ calculations are shown in Figure~\ref{fig:gsvsbs-phi}.  These are shown for both good and poor solvent conditions as indicated, plotted as a function of crowder densities and for a fixed wall attraction strength ($\epsilon_{mw}=10.0$) 
From Figure~\ref{fig:gsvsbs-phi}, it can be seen that for poor solvents, the polymer adapts an extended conformation on the confining spherical surface, if the  attractive wall strength is large and the crowder density low.  However, as the crowder density is increased, the polymer configuration undergoes a transition from more extended configurations to a collapsed configuration. This is in contrast to good solvent conditions, for which polymer conformations are relatively insensitive to crowder density, as shown in Figure~\ref{fig:gsvsbs-phi}(a). 

The calculated values of $R_g$ and $f_{ad}$ as a function of crowder densities in Figure~\ref{fig:gsvsbs-phi}(c-d) quantify the transition from extended to collapsed polymer conformations in poor solvent conditions and at high crowder densities. We contrast these results to the case for low crowder density in Figure \ref{fig:snaps-diffE13}(a) where, as the attractive wall strength is increased, a first transition from collapsed to extended conformation (AC to AE phase) in poor solvent is observed.  While that transition was driven by the attractive wall strength, the transition from extended to collapsed shown in Figure~\ref{fig:gsvsbs-phi} (b) in the poor solvent case is driven by crowder density. 

\begin{figure}
\centering
\includegraphics[width=\linewidth]{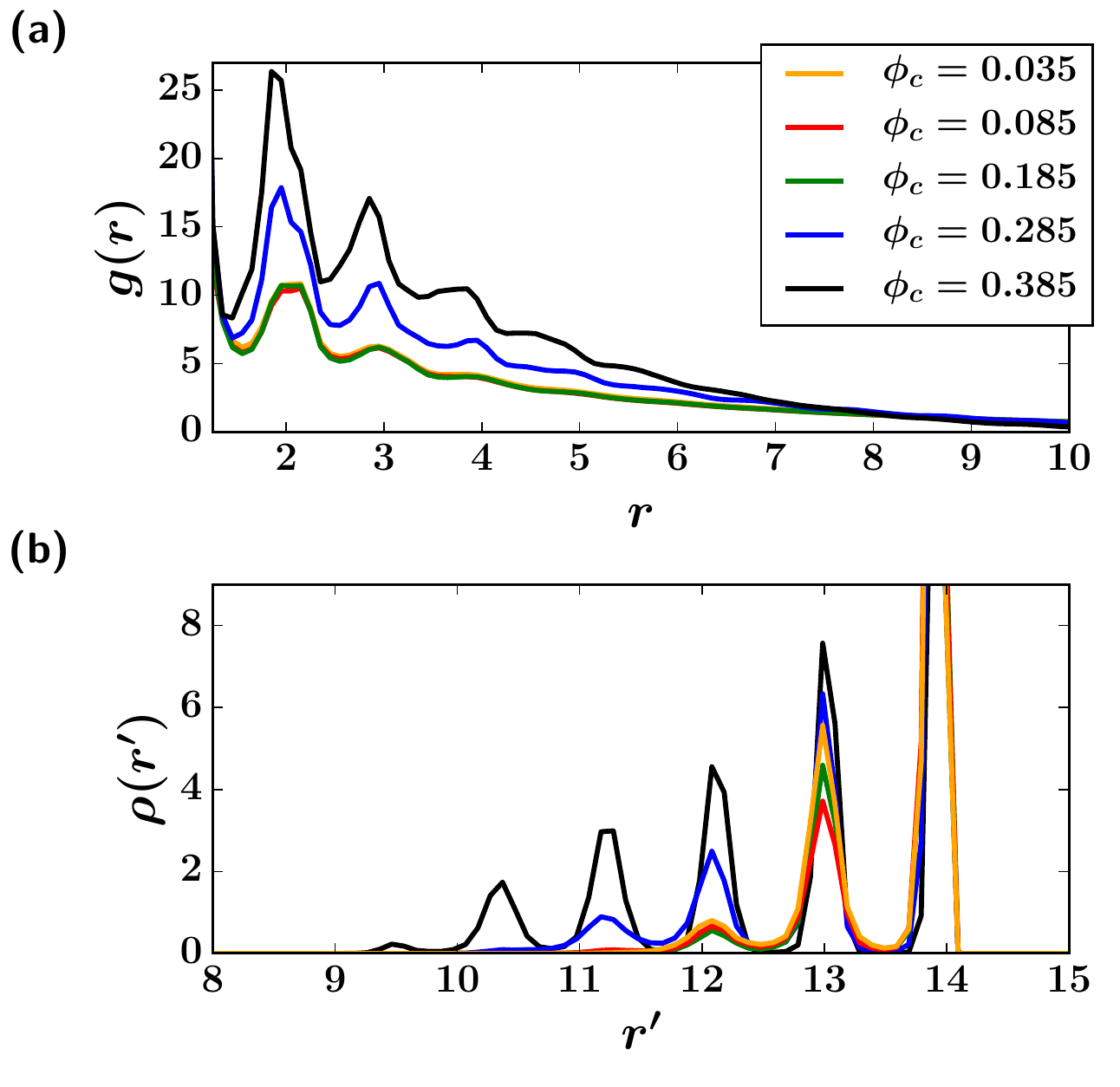}
\caption{For poor solvent condition and $\epsilon_{mw} = 10.0$ \textbf{(a)} the radial distribution function $g(r)$, \textbf{(b)}  the density function $\rho(r^{\prime})$ for $\phi_c = 0.035, 0.085, 0.185, 0.285, 0.385$.}
\label{fig:gofr-dfun}
\end{figure}

To investigate structure in the conformation of the extended polymer at high wall interaction strength, we calculate the radial distribution function $g(r)$. For fixed monomer-wall interaction ($\epsilon_{mw} = 10.0$), as $\phi_c$ is increased, the order in the structure increases (see Figure~\ref{fig:gofr-dfun} (a)). To understand the positioning of the polymer relative to the wall, we plot the normalized density, $\rho(r')$, of the polymer vs. the radial distance $r'$ (see Figure~\ref{fig:gofr-dfun} (b)) from the center of the sphere. To calculate $\rho(r')$, we split the spherical volume in a large number of thin shells  and count the number of monomers in each shell, normalizing this with respect to the average density obtained if the monomers were distributed uniformly throughout the volume. This density distribution $\rho(r')$ gives us insights into the layering of the polymer upon increasing crowder density. 

Our results suggest that, as the crowder density is increased, there is an emergence of  layering with respect to the surface. For lower crowder density, there are fewer layers. As  we increase the crowder density in the system, more layers begin to appear, as seen in  Figure~\ref{fig:gofr-dfun} (b)). This can be understood in the following way. In poor solvent conditions, as we increase the number of crowder particles in the system, the polymer tries to collapse in order to avoid the exposure to the crowder particles. However, since monomers are attracted towards the wall, a fully globular conformation cannot be sustained. This leads to a partial layering near the wall and hence to the peaks in the density plot. The emergence of internal structure can also be captured via time-averaged contact maps shown in Figure~S4 (Supplementary Information). The contact maps show that as the crowder density increases, the number of contacts increase in general, suggesting an increase in local density. There are also significant  contacts between monomers which are  far apart along the sequence, consistent with the visual identification of the collapse of the polymer with increasing crowder density. 

Finally, to understand the contribution of polymer degrees of freedom to the entropy of the system, we simulated un-connected monomers instead of linking them into a $N = 400$ polymer. Our results, shown in Figure~S5 (Supplementary Information), are the following: For poor solvents, at all crowder densities less than about 0.3, all the beads stick independently to the surface and are spread out on it, as evidenced in the plots of $R_g$, $b$, $f_{ad}$ and $H$. For higher crowder densities, the particles prefer to form a denser, yet still 2-dimensional adhered layer, with traces of crystalline order. For good solvents, essentially the same behaviour is seen, except that the clustering at high crowder densities is absent. As can be seen, the configurations for isolated beads are substantially different from those in the polymer case, a feature that can only follow from the difference between bonded and non-bonded monomers. 

\subsection{A crumpling transition induced by crowders}
\begin{figure}
\includegraphics[width=\columnwidth]{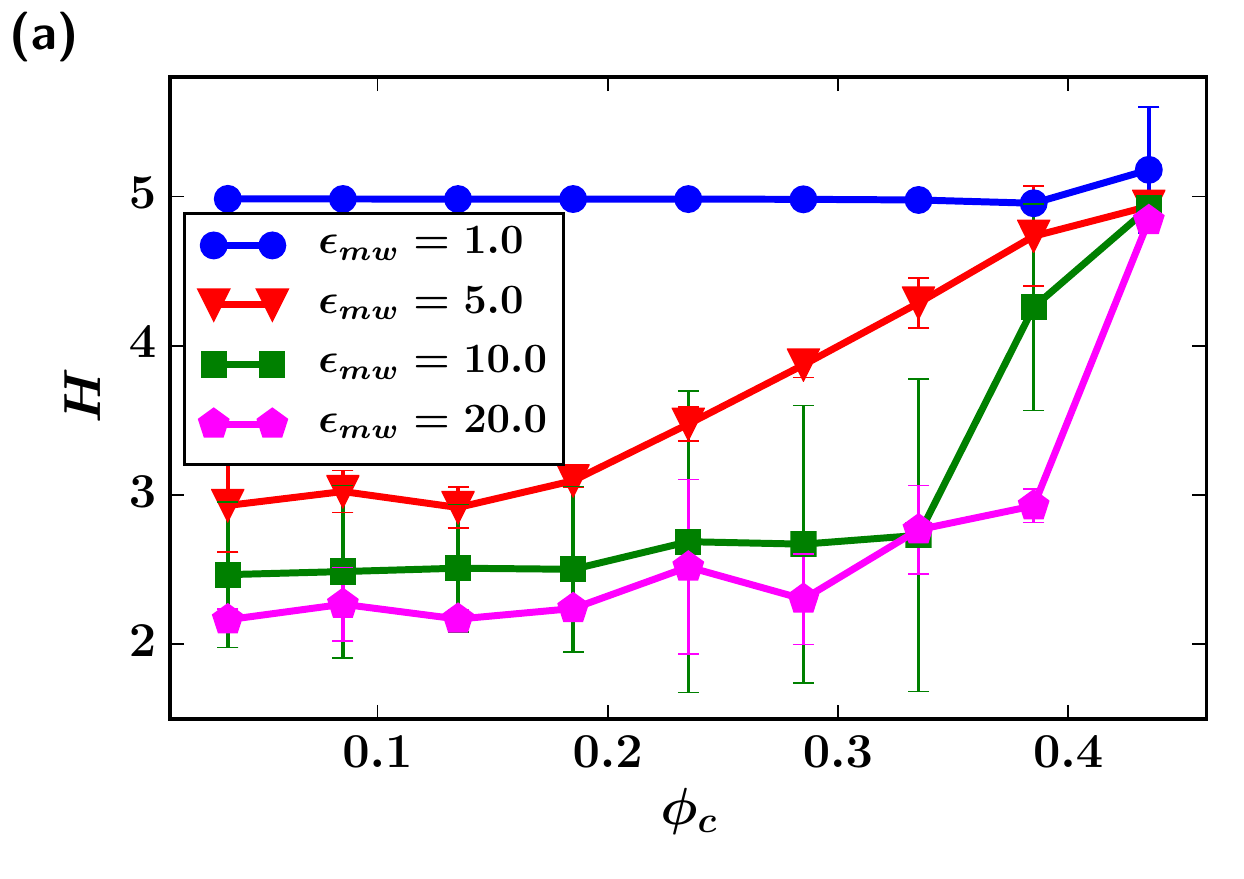}
\includegraphics[width=\columnwidth]{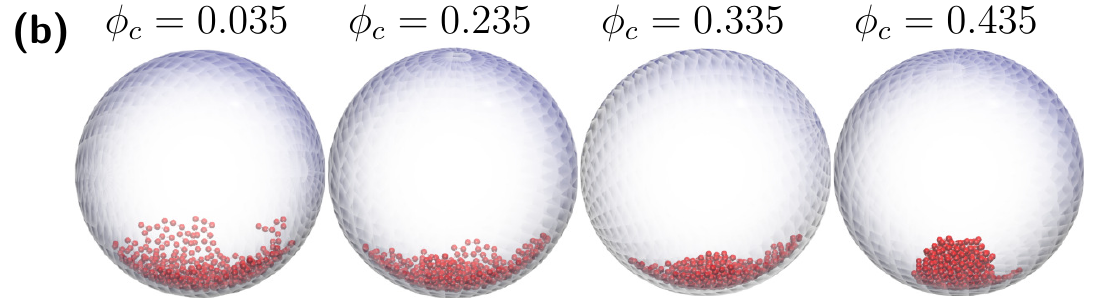}
\caption{\textbf{(a)} The maximum height of the polymer stack along the radial direction. The height of the stack decreases as the strength of the attraction increases and more and more monomers are recruited along the wall of the confinement. \textbf{(b)} The snapshots of the polymer conformation in poor solvent condition for the different crowder densities.}
\label{fig:height}
\end{figure}

To better characterize the ``crumpling'' of the polymer under the combination of poor solvent conditions and high crowder densities, we plotted the maximal height of the polymer from the surface as a function of crowder density, for all wall interaction strengths (see  Figure~\ref{fig:height}(a)). 

Some intuition for the maximal height can be obtained from side-views of the system as a function of crowder density, as shown in Figure~\ref{fig:height}(b). At low crowder densities, increasing the strength of the attractive interactions between monomers and wall results in the polymer spreading on the surface. This is evident in the height profile at the smallest crowder density considered ($\phi_c=0.035$). The highest height is achieved for the smallest wall attraction strength, where the polymer assumes a collapsed conformation. When the attractive strength of the wall is small $(\epsilon_{mw} = 1.0)$, the height function is largely independent of the crowder density except at the largest crowder density, where the compactness of the polymer increases. However, for other attractive wall strengths, the dependence of the height function on the crowder density is more complex. For an attractive wall strength of $\epsilon_{mw} = 5.0$, the initial configuration of the polymer is more extended (as can be seen from the difference in heights even at the smallest crowder density). However, the extended-to-collapsed transition is initiated at a lower density than in the case of $\epsilon_{mw} = 1.0$. For higher wall interactions, the extended-to-collapsed transition occurs at higher crowder densities. 

Our results suggest the following: The polymer in poor solvent conditions near an attractive wall  experiences two competing forces. These favour, independently, compaction and spreading on the wall. The propensity to collapse would lead to a higher height function. This is primarily driven by crowder particles. The propensity of the polymer to spread along the surface arises from its attractive interactions with the surface. The interplay between these  determines the eventual conformational landscape of the polymers at the interface. For low attractive wall strength $\epsilon_{mw}=5.0$, a smaller threshold crowder density is required for collapse (for $\phi_c < 0.235$). We thus see an increase in the height function. As the strength of the  attraction towards the wall is increased, the threshold crowder density after which the polymer starts to collapse moves towards higher crowder density, as seen for $\epsilon_{mw} =10.0, 20.0$. At the highest crowder densities ($\phi_c=0.435$), irrespective of wall interactions, the polymer under poor solvent conditions is predominantly in a collapsed state, as evident  in the height function. 

Snapshots of the collapsed polymer in poor solvent conditions under lowest and highest crowder densities are shown in Figure~S6(a,b) (Supplementary Information). The figure shows two collapsed structures obtained under two very different conditions. From the images, we can already see internal structure in these at high crowder density. To quantify the emergence of internal structure in the collapsed conformation at high crowder densities,  we calculated the pair distribution function for the two collapsed phases.  These results are shown in Figure~S6(c).  The presence of strong peaks in the distribution function, for the case of high crowder density,  clearly indicates increased structure in the collapsed conformations compared to the case for lower crowder density. 

To further quantify  local structure, we calculate the average number of neighbours for each monomer and plot it is a function of monomers in Figure~S7.  From the results, it can be seen that the average number of neighbours, barring the monomers on the surface of the collapsed conformation, is around $9$, which is very different from the low crowder density case. This suggests that the polymer assumes a  more ordered collapsed phase at high crowder density and a more amorphous collapsed phase at low crowder density.

\section{Discussion and Conclusion}

In this paper, we used molecular dynamics simulations to describe the conformational landscape of a confined polymer in the presence of attractive walls as well as crowder particles. We varied solvent quality across good and poor solvent conditions. We used a $400-$monomer long polymer, exploring the parameter space of the density of  crowder particles, the strength of wall interactions and the quality of the solvent.  

The conformation of a polymer on a surface has been the subject of a number of earlier studies \cite{rajesh2002adsorption, bachmann2005conformational, plascak2017solvent, martins2018adsorption}. The nature of polymer configurations depends on the curvature of the surface \cite{arkin2012structural, bachmann2005conformational, moddel2014adsorption}, solvent quality \cite{reddy2010solvent}, boundary conditions, and the interaction energy between the surface and the polymer. 

Earlier Monte Carlo studies \cite{moddel2014adsorption} showed that polymers display a number of shape transitions near an attractive planar surface. This work suggested a pseudo phase-diagram in the $\epsilon_{mw}-T$ plane. In this phase diagram, at low temperatures, the polymer assumes a layered adsorbed crystalline shape. At higher temperatures, the polymer is desorbed if the  surface attraction is small, while the polymer remains adsorbed for high surface attraction. Similar shape transitions are seen for polymers on curved surfaces \cite{arkin2012structural}. 

As we show in this paper, the situation is considerably more complex when the polymer is confined, thereby reducing its allowed fluctuations. We showed that for poor solvents, as the attractive interaction strength between the wall and the polymer was increased, the polymer exhibited a shape transition from an adsorbed globule (AG) to an adsorbed layered (AL) phase.  For  intermediate values of $\epsilon_{mw}$, as the number of crowder particles was increased, polymer  configurations changed from the AL state to an adsorbed collapsed (AC) state.  In good solvents, at low surface attraction, the polymer did not completely adsorb on the surface.  As the surface attraction was increased the polymer gradually adsorbed on the attractive surface. 
 
The curvature of the confining sphere plays an important role in the selection of conformations. We observed layered structures in proximity to the wall, as seen from the difference in number of peaks (suggesting layers of stacked monomers) when the radius of confining sphere was doubled (see Figure~S8). As the radius of the confining sphere is increased, the number of secondary peaks decreases. For the larger sphere, all the monomers are adsorbed on the surface at low $\phi_c$. As the density of crowders is increased, a second layer appears. For the smaller system, there are multiple layers present even at the smaller density of crowders. These results suggest that the confinement and the relative ratio of confining sphere radius and length of the polymer can significantly alter the conformational landscape of the polymers.  In particular, smaller confinement radii can led to more substantial layering, presumably because the increased curvature of the smaller sphere frustrates the formation of the adsorbed monolayer that is a feature of the sphere with larger radius. These features can all be tuned by the density of crowders, providing a second axis to adjust the properties of the adsorbed state.

In this study, in addition to the interacting surfaces and solvent condition, we discussed the role of another critical variable associated to crowding of the polymer chain by other monomeric particles. The quantity $\phi_c$ was found to play an important role in determining the conformation of the polymer. In good solvent conditions, we found that the crowder particle density did not affect the conformations substantially.  In poor solvent conditions however, for higher levels of wall attraction, an increase in number of crowder particles leads to a more ordered structure. We suggest that this may be a general feature of crowded confined polymers in a poor solvent, when attracted by a wall, and that the ability to adjust solvent quality as well as crowder concentration may be key to stabilizing a polymer in the vicinity of a confining wall.

The addition of crowder particles adds additional complexity but presumably provides a better representation of a large number of biological situations. Indeed, the combination of confinement, crowding and wall interactions should be generic to a number of polymer systems of biological and pharmacological relevance. More detailed studies of these regimes, keeping specific systems in mind, and including the size effects of the crowders under these conditions should be fruitful.
\onecolumngrid
\begin{center}
\begin{figure}
\includegraphics[width=0.8\linewidth]{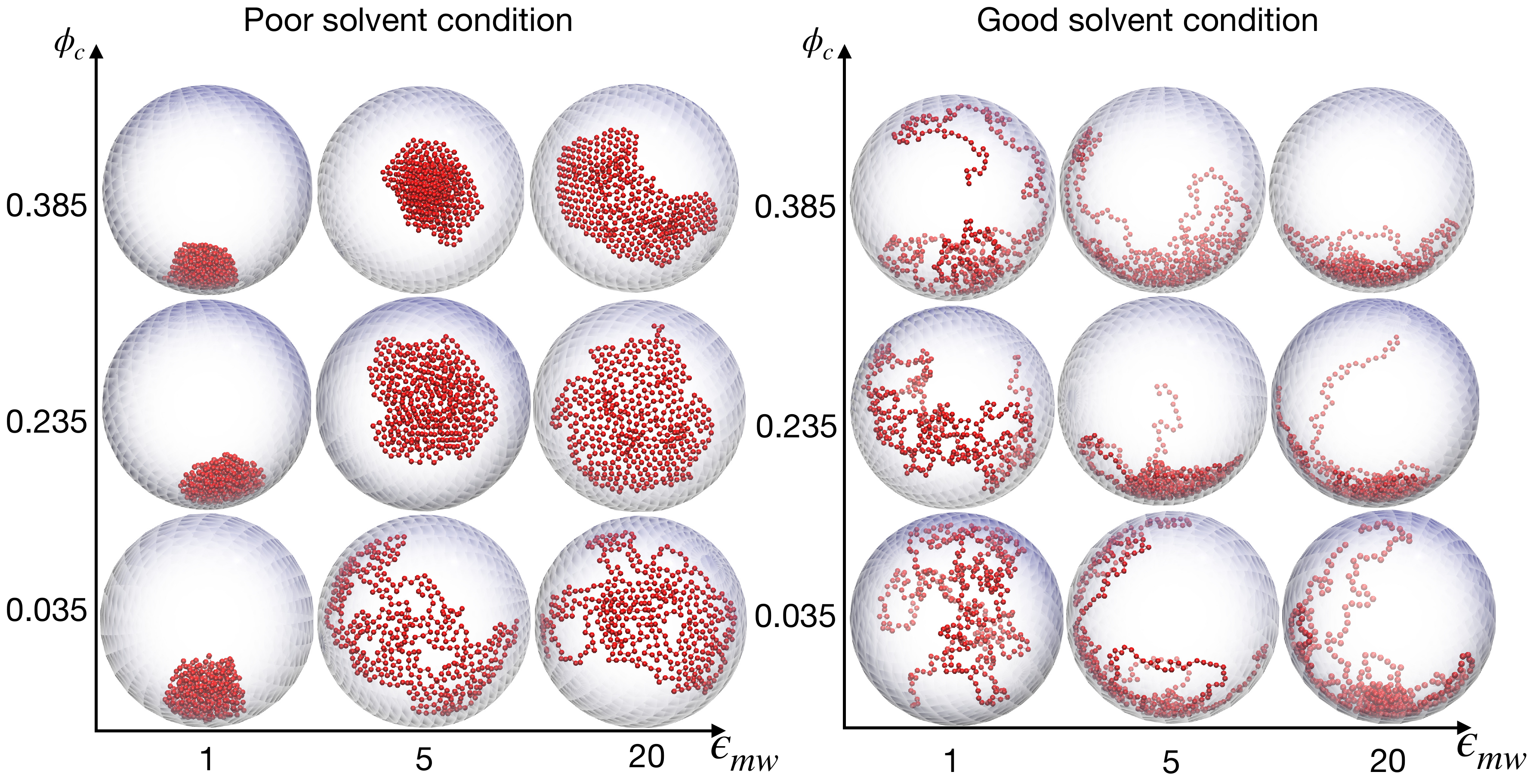}
\caption{Conformations of polymer under different crowder densities and attractive wall strengths for the poor and good solvent condition in $\phi_c-\epsilon_{mw}$ space.}
\label{fig:conf-phase}
\end{figure}
\end{center}
\twocolumngrid

\section{SUPPLEMENTARY INFORMATION}
See supplementary information for additional analyses and interaction parameters.

%
\end{document}


\title{\Huge \underline{Supplementary Information}\\ 
{\huge {Confined crowded polymers near attractive surfaces }} \\
}
\author[1,2]{Kamal Tripathi} 
\author[1,2,3]{Gautam I. Menon} 
\author[1,2]{Satyavani Vemparala} 
\affil[1]{The Institute of Mathematical Sciences, C.I.T. Campus, Taramani, Chennai 600113, India}
\affil[2]{Homi Bhabha National Institute, Training School Complex, Anushakti Nagar, Mumbai 400094, India}
\affil[3]{Department of Physics, Ashoka University, Plot No. 2, Rajiv Gandhi Education City, National Capital Region, Sonepat 131029, India}
\date{}
\maketitle
\newpage
\section{\textcolor{black}{Relative change in $R_g$ values, as a function of crowder density}}
\begin{figure}[H]
\centering
\includegraphics[width = 0.6\linewidth]{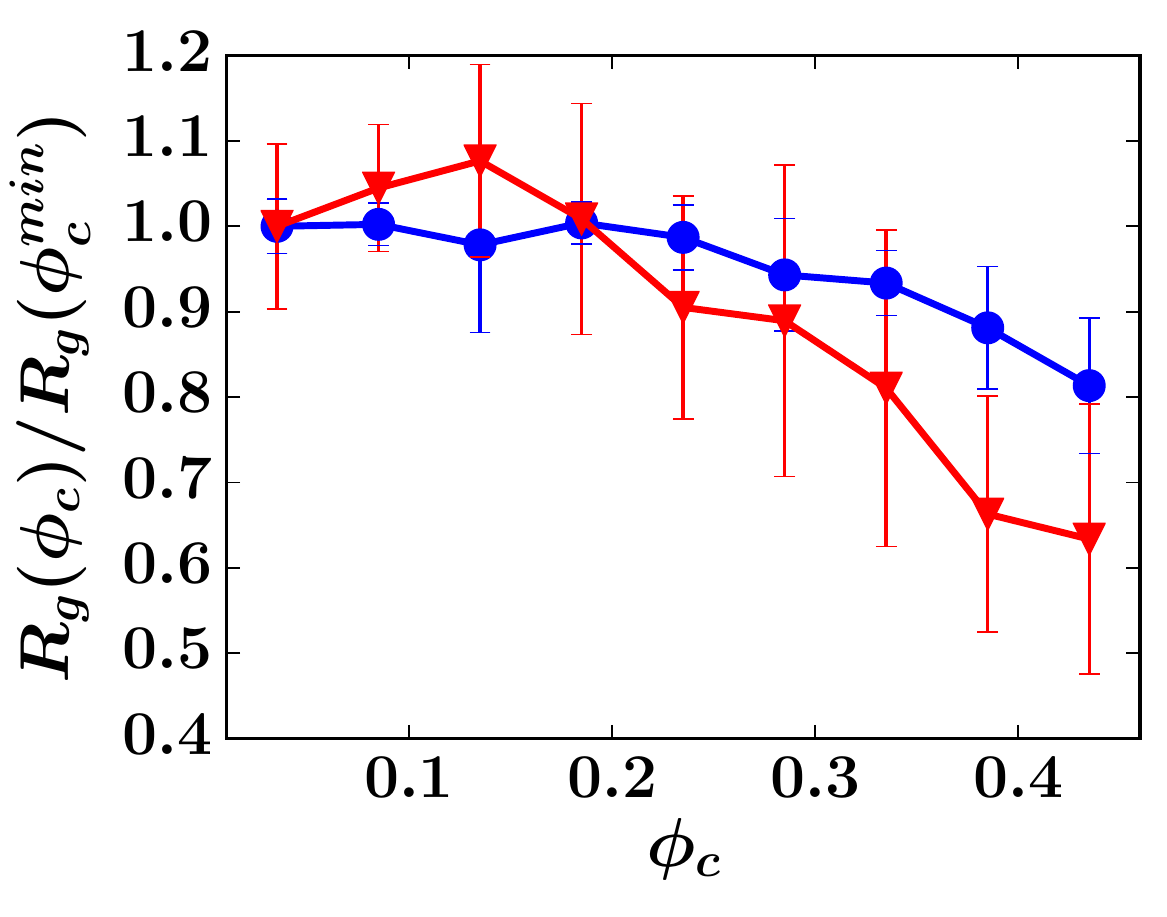}
\caption{Relative change in the $R_g$ values, as a function of crowder density, for the unconfined (red) and confined (black) cases. The values are computed relative to the case of lowest crowder density considered in the study ($\phi_c^{min}=0.035$). Each data point is averaged over 10 different initial conditions.
\label{fig:rg-scaled}}
\end{figure}

\section{Effect of degree of polymerization in good solvent condition}
\begin{figure}[H]
\centering
\includegraphics[width=0.45\columnwidth]{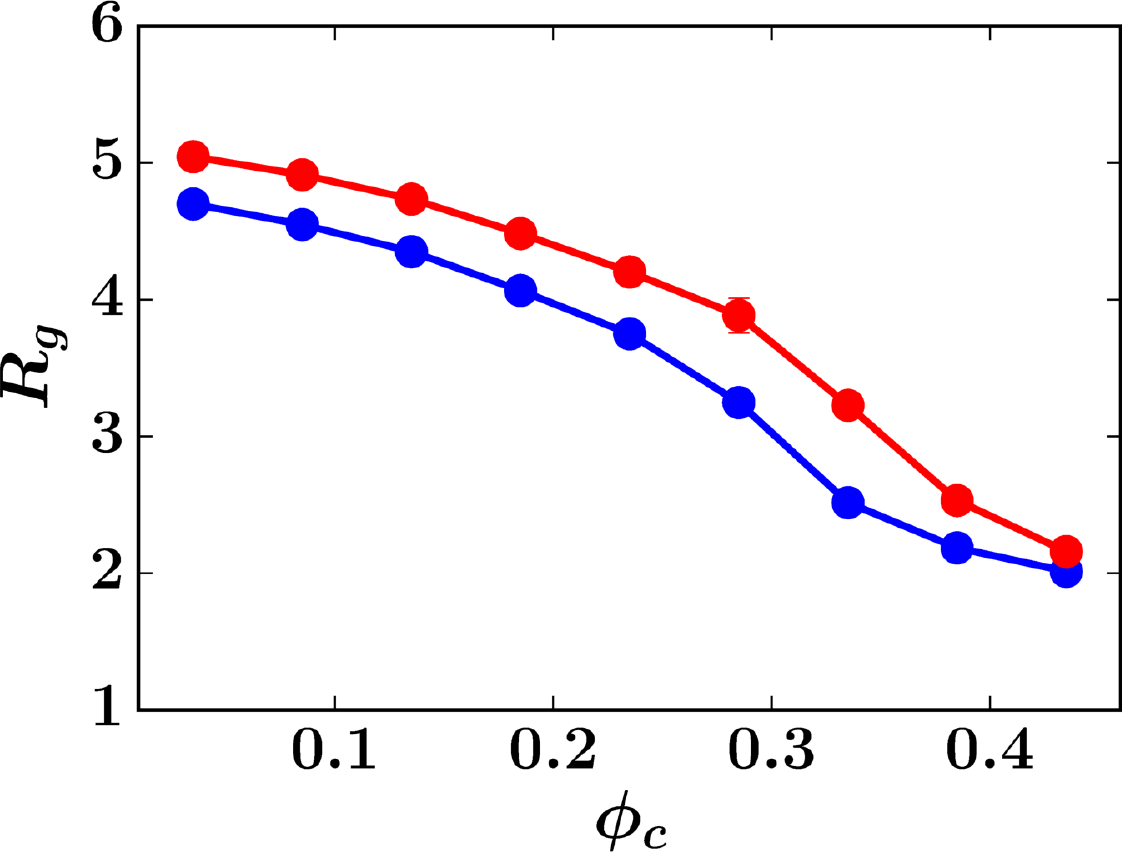}
\includegraphics[width=0.45\columnwidth]{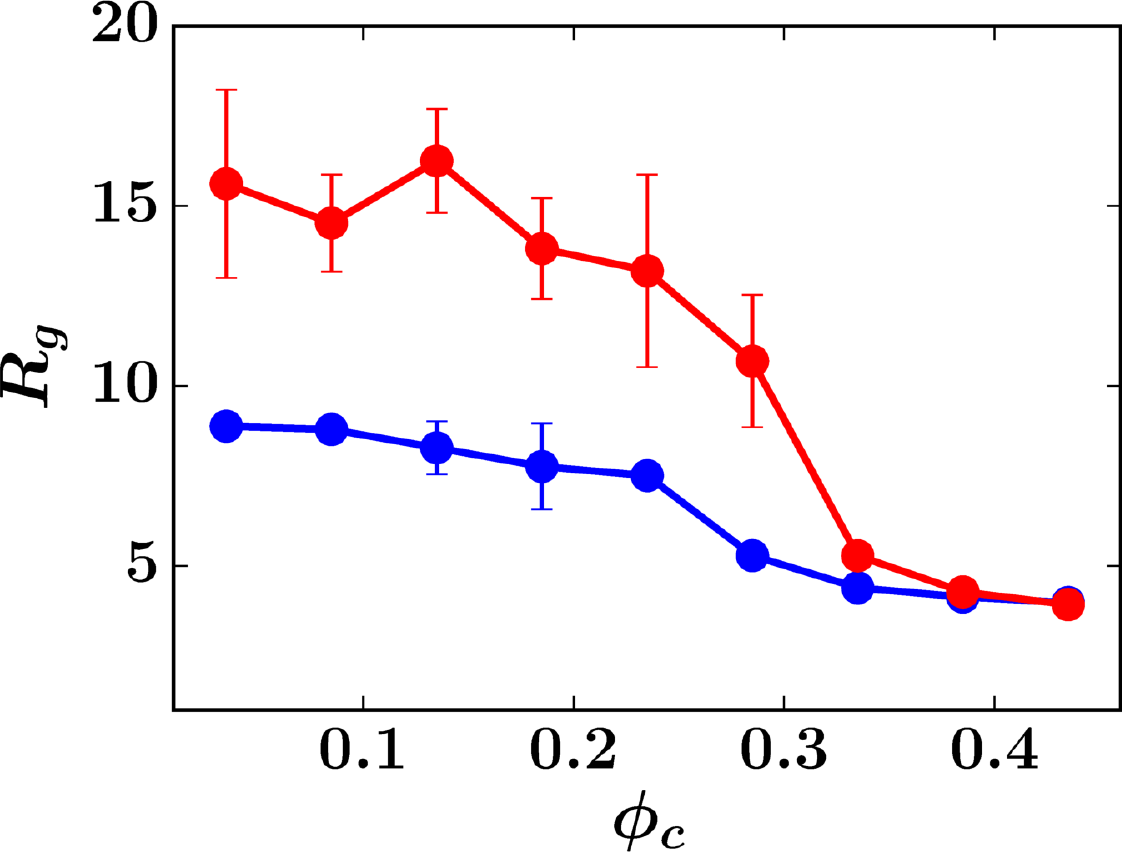}
\caption{The radius of gyration of the polymer in a good solvent condition under confinement with repulsive walls (blue) and in the absence of confinement (red) at different crowder densities with $\epsilon_{mc} = 5.0$, \textbf{(a)} for the polymer with degree of polymerization $50$ \textbf{(b)} for the polymer with degree of polymerization $400$.}
\label{fig:Rg-DegPoly}
\end{figure}

\section{Good solvent condition with $\epsilon_{mc} = 5.0$}
\begin{figure}[H]
\includegraphics[width=\columnwidth]{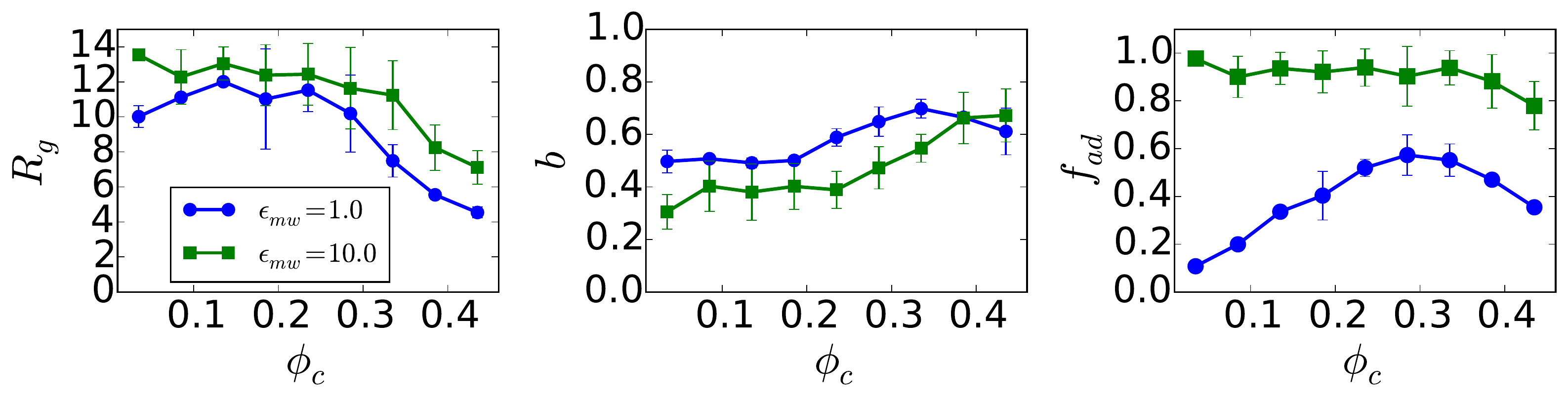}
\caption{The variation of \textbf{(a)} radius of gyration $R_g$, \textbf{(b)} asphericity $b$ and \textbf{(c)} the fraction of adsorbed monomers $f_{ad}$ with $\phi_c$ for $\epsilon_{mc} = 5.0$ keeping $\epsilon_{mm}$ and $\epsilon_{cc}$ same as earlier for different $\epsilon_{mw} = 1.0$ and $10.0$}
\label{fig:gsaw-Emc-5}
\end{figure}

\section{Contact maps as a function of crowder densities for poor solvent conditions}
\begin{figure}[H]
\centering
\includegraphics[width=\columnwidth]{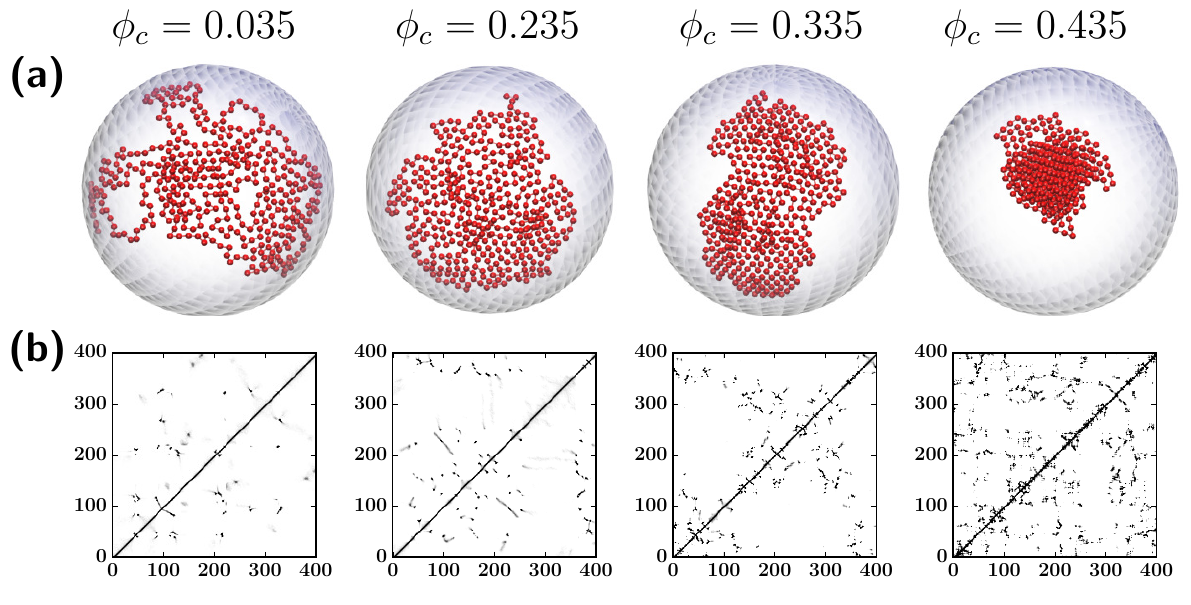}
\caption{For poor solvent condition and $\epsilon_{mw} = 20.0$: \textbf{(a)} Snapshots of the system for different values of $\phi_c$, and \textbf{(b)} Corresponding contact maps averaged over the final  $5 \times 10^6$ time steps.}
\label{fig:contmap-phi}
\end{figure}

\section{Effect of bonded-interaction in poor and good solvent condition}
\begin{figure}[H]
\includegraphics[width=\columnwidth]{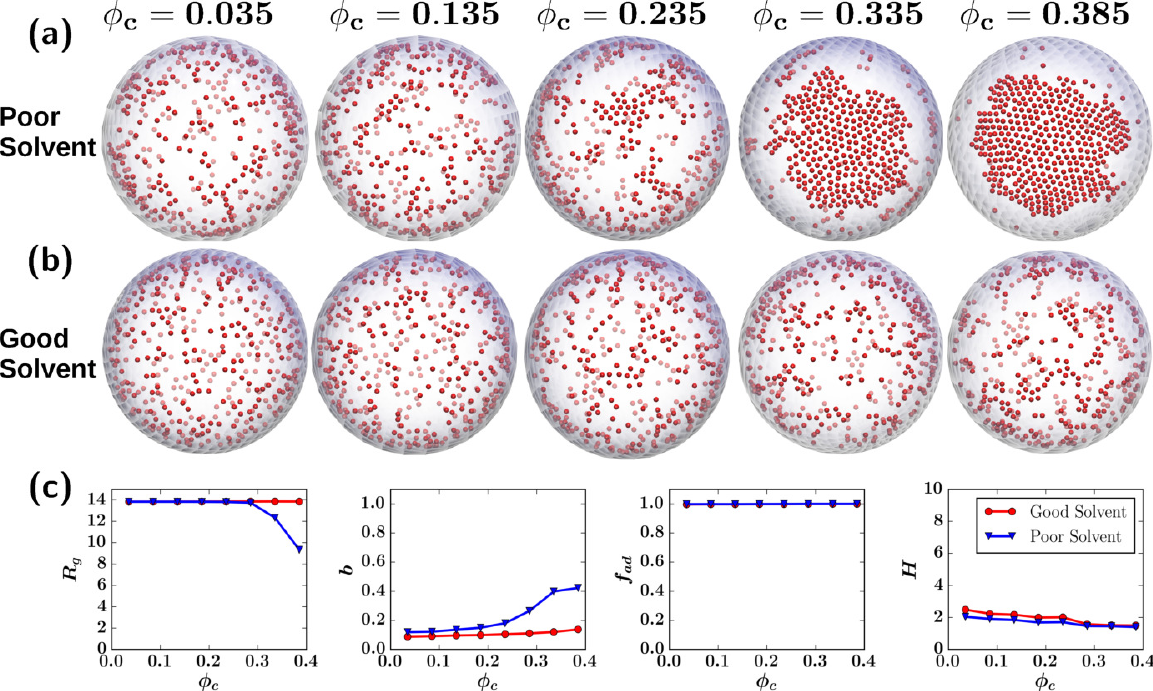}
\caption{Conformation of monomers in \textbf{(a)} in poor solvent condition and in \textbf{(b)} good solvent condition. Also shown are plots of \textbf{(c)} the radius of gyration $Rg$, asphericity $b$, fraction of adsorbed monomers $f_{ad}$ and the height $H$, from left to right, for different $phi_c$ values, for both  poor and good solvent condition. The results in the plot are from a single long simulation with $\epsilon_{mc} = 5.0$.} 
\label{fig:loosemono}
\end{figure}

\section{Radial distribution function for low and high crowder density}
\begin{figure}[H]
\centering
\includegraphics[width=\columnwidth]{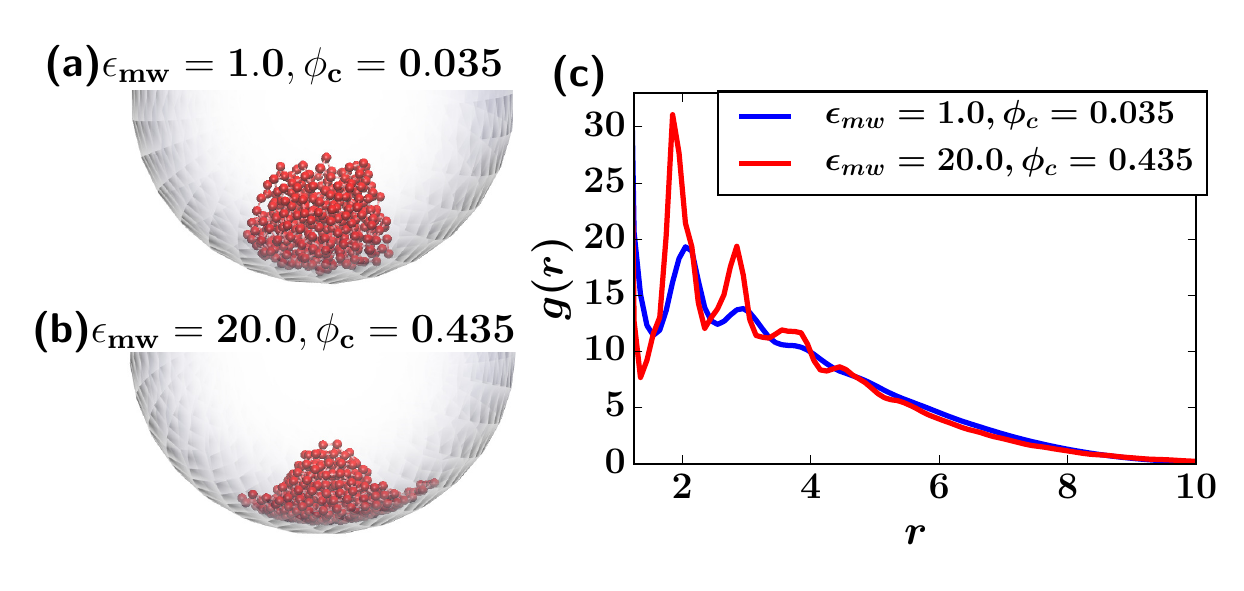}
\caption{ Snapshots of the system in a poor solvent for \textbf{(a)} $\epsilon_{mw} = 1.0, \phi_c = 0.035$,  \textbf{(b)} $\epsilon_{mw} = 20.0, \phi_c = 0.435$ and \textbf{(c)} the pair radial distribution function of polymer for parameters of \textbf{(a)} and \textbf{(b)} . For lower wall attraction $\epsilon_{mw}$ and lower density value $\phi_c = 0.035$, the plot suggests less order while for the higher wall attraction $\epsilon_{mw} = 20.0$ and higher crowder density $\phi_c = 0.435$, $g(r)$ shows many peaks indicating more order in the system.}
\label{fig:gofr}
\end{figure}

\section{Average number of neighbours}
\begin{figure}[H]
\includegraphics[width = \linewidth]{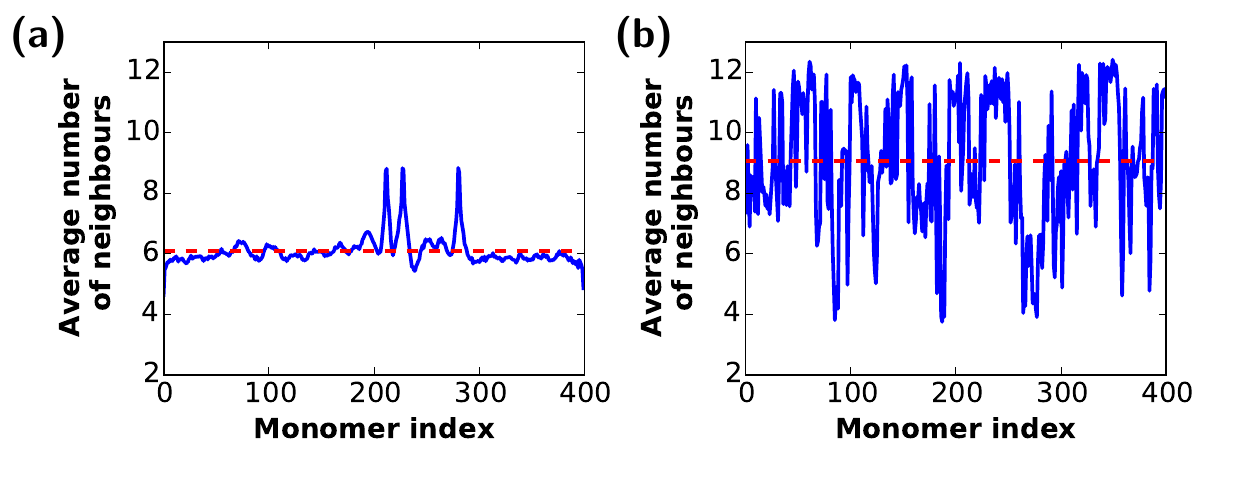}
\caption{Average number of neighbours for \textbf{(a)} $\epsilon_{mw} = 1.0$, $\phi_c = 0.035$ and \textbf{(b)} $\epsilon_{mw} = 20.0$, $\phi_c = 0.435$. The average number of neighbours for higher density $(\phi_c = 0.435)$  and higher wall attraction ($\epsilon_{mw} = 20.0$) system is about $9$ which is higher than that for lower wall attraction ($\epsilon_{mw} = 1.0$) and lower crowder density $(\phi_c = 0.035)$ system. This suggests that polymer is more ordered in the higher crowder density system than for the  lower density system.}
\label{fig:av-neigh}
\end{figure}

\section{Effect of curvature on the conformation of polymer in poor solvent condition}
\begin{figure}[H]
\includegraphics[width=\columnwidth]{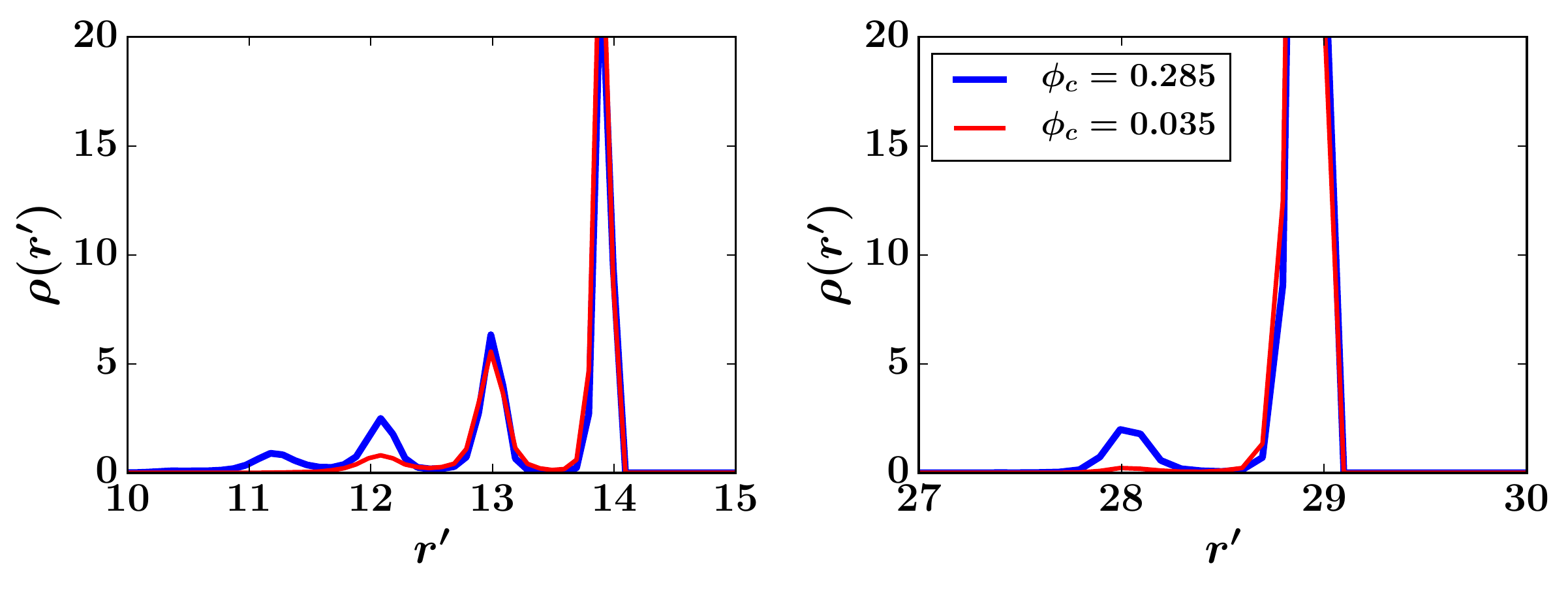}
\caption{The density function of the monomers in poor solvent condition confined within attractive walls \textbf{(a)} for the sphere of radius $R = 15.0$ \textbf{(b)} for the sphere of radius $R = 30.0$.}
\label{fig:rhoR-diffR}
\end{figure}
In order to understand the effect of the curvature on the conformation, we simulated the system of polymer of same length in a sphere with larger radius $R = 30.0$ with the total volume fraction from $0.035$ to $0.435$ for $\epsilon_{mw} = 10.0$. It is seen that increasing the radius of the sphere leads to a higher fraction of adsorption of the monomers. The density function describes the normalized number of particles at a given distance $r'$ from the center of the confinement. The density function plots for monomers indicate that for the system with $R = 15.0$ and $\phi_c = 0.035$, there are two layers of monomers. Increasing the $\phi_c$ to $0.285$ leads to multiple peaks (see Figure~\ref{fig:rhoR-diffR}(a)) which is an indication of the onset of collapse of the polymer to a globular conformation. In the case of the system with larger radius $R = 30.0$ and for $\phi_c = 0.035$, all the monomers are adsorbed on the surface while increase in $\phi_c$ value to $0.285$, some of the monomers associate to the top of the monomer layer, indicated by the emergence of the second peak in the density function plot (see Figure~\ref{fig:rhoR-diffR}(b)) .

\section{Simulation Parameters}
\begin{table}[H]
\setlength{\tabcolsep}{14pt}
\begin{center}
\begin{tabular}{ |c|c|c|c|c|c|c| } 
\hline
\multicolumn{1}{|c|}{} & \multicolumn{3}{c|}{\textbf{Good solvent}} & \multicolumn{3}{c|}{\textbf{Poor solvent}}\\
\hline
$\mathbf{ i-j}$ & $\mathbf{\epsilon_{ij}}$ & $\mathbf{\sigma_{ij}}$ & $\mathbf{r_c}$ & $\mathbf{\epsilon_{ij}}$ & $\mathbf{\sigma_{ij}}$ & $\mathbf{r_c}$\\
\hline
 m-m & 1.00 & 1.00 & $(2)^{1/6}$ & 1.00 & 1.00 & 2.5\\ 
 c-c & 1.00 & 1.00 & $(2)^{1/6}$ & 1.00 & 1.00 & $(2)^{1/6}$\\ 
 m-c & 0.50 & 1.00 & $(2)^{1/6}$ & 5.00 & 1.00 & $(2)^{1/6}$\\ 
\hline
\end{tabular}
\end{center}
\caption{Table of parameters for LJ potential for good and poor solvent conditions. Different values of $r_c$ determine the solvent condition. To delineate the effect of poor solvent conditions, a higher value of $\epsilon_{ij}$ was used between monomers and crowder particles, though the potential is soft-core repulsion for good and bad solvent cases.  }
\label{tab:LJpar}
\end{table}

\clearpage